\documentclass[amssymb,aps,twocolumn,showpacs,nofootinbib,prx,longbibliography]{revtex4-1}
\usepackage[]{graphicx}
\usepackage{amsmath}
\usepackage[colorlinks=true,urlcolor=blue]{hyperref}
\usepackage{color}

\newcommand{\ket}[1]{{\left\vert{#1}\right\rangle}}

\newcommand{\proj}{\mathbb{P}}
\newcommand{\set}[1]{\textrm{\textbf{#1}}}
\newcommand{\op}{\mathcal}
\renewcommand{\vec}{\mathbf}
\newcommand{\tr}{\textrm{tr}}
\newcommand{\ave}{\overline}

\begin{document}

\title{Randomized benchmarking with restricted gate sets}
\author{Winton G. Brown}
\email{Winton.Brown@ngc.com}
\author{Bryan Eastin}
\affiliation{Northrop Grumman Corporation }

\date{\today}

\begin{abstract}
Standard randomized benchmarking protocols entail sampling from a unitary 2 design, which is not always practical.  In this article we examine randomized benchmarking protocols based on subgroups of the Clifford group that are not unitary 2 designs.  We introduce a general method for analyzing such protocols and subsequently apply it to two subgroups, the group generated by controlled-NOT, Hadamard, and Pauli gates and that generated by only controlled-NOT and Pauli gates.  In both cases the error probability can be estimated to within a factor of two or less where the factor can be arranged to be conservative and to decay exponentially in the number of qubits.  For randomized benchmarking of logical qubits even better accuracy will typically be obtained.  Thus, we show that sampling a distribution which is close to a unitary 2 design, although sufficient, is not necessary for randomized benchmarking to high accuracy. 
\end{abstract}

\pacs{}

\maketitle

\section{Introduction}
Randomized benchmarking is widely used for characterizing the performance of quantum information processing devices~\cite{BWC+11,GMT+12,CGT+09,BKM+14,RLL09,OCNP10, XLM+15, MLS+15}.
Standard randomized benchmarking protocols require the capability to sample from a unitary 2 design~\cite{KLR+08,DCEL09, MGE11, MGE12}, for which the Clifford group, or certain subgroups of the Clifford group which are nonetheless unitary 2 designs, are usually used.
Sampling from a 2 design can be challenging, however, particularly for the case of randomized benchmarking of logical qubits, where the set of high-fidelity logical gates that can be implemented in a straightforward manner is invariably restricted. 

For many quantum codes, techniques such as state injection and distillation are required to implement some generators of the logical Clifford group.  Clifford gates incorporating such generators suffer from much higher overhead and probability of error.
Logical qubits are likely to be in short supply for some time, so overhead is a significant concern for near-term demonstrations of randomized benchmarking on logical qubits.  Furthermore, incorporating logical Clifford gates with low utility and poor performance in logical randomized benchmarking is undesirable as it results in an overly pessimistic assessment of the logical gate set.  The latter issue might be resolved by benchmarking individual logical gates via interleaved randomized benchmarking~\cite{MGJ+12} except that this technique has been found to perform poorly for gate sets with wildly varying error probabilities~\cite{ECMG14}.  Faced with these issues, an obvious question to ask is whether the offending gates are truly necessary for randomized benchmarking.  We show here that in many cases they are not.

The fundamentals of randomized benchmarking are briefly reviewed in Section~\ref{sec:randomizedBenchmarking}.  Section~\ref{sec:CliffordSubgroupTwirling} introduces randomized benchmarking using subgroups of the Clifford group and analyzes the performance in the aforementioned cases of interest.  Concluding remarks appear in Section~\ref{sec:conclusion}.

\section{Randomized Benchmarking\label{sec:randomizedBenchmarking}}

\subsection{Setting and Assumptions}

 The goal of randomized benchmarking is to determine the average fidelity of the error channels of a group of unitary gates of interest $\set{G}$.  For the benchmarking procedure to accurately return this value, the set of error channels associated with the gates must satisfy the following properties.
\begin{enumerate}
\item{{\bf Markovianity:} For any target $U \in \set{G}$ the true evolution of the quantum device can be expressed as a completely positive map,
$\op{E}_{U} \circ \hat{U}$, where 
$$\op{E}_{U}\circ \hat{U}(\rho) = \sum_i E_i U\rho U^\dagger E_i^\dagger.$$

}
\item{{\bf Weak gate dependence:}  The error channel does not depend strongly on the gate applied.
That is, if $\op{E}_{U}$ is the error channel associated with $U$ and $\op{E} = \frac{1}{|\set{G}|} \sum_{U \in \set{G}} \op{E}_{U}$ then
$$ \|\op{E} - \op{E}_{U} \|_1 \ll 1, $$
where the norm used is the induced operator 1-norm.
}
\end{enumerate}

Recently, some complications with regard to weak gate dependence have been noted~\cite{PRY+17,Wal17}.  For simplicity, we further assume that the error channel is completely independent of the gate applied.

\subsection{Twirling}

The basic approach utilized by randomized benchmarking is to symmetrize a quantum channel, $\op{E}$, with respect to a group of unitary gates, $\set{G}$, by performing a random gate from $\set{G}$ on the input and its inverse on the output of the channel.  Once symmetrized, the channel can be described by a small number of parameters which may then be determined efficiently by a simplified tomography procedure.

Specifically, performing a random gate, $U$, sampled from $\set{G}$, on the input of the channel, $\op{E},$ while performing its  inverse, $U^\dagger$, on the output results in the following channel:
$$\tilde{\op{E}}_\set{G} =\mathbb{E}_{U\sim\set{G}}  \left[\hat{U}^\dagger\circ \op{E} \circ \hat{U}  \right],$$ 
where $\mathbb{E}_{U\sim \set{G}}$ represents an average over all $U$ sampled uniformly from $\set{G}$.
The channel $\tilde{\op{E}}_\set{G}$ is symmetrized in that it is invariant under conjugation by elements of the group $\set{G}$, that is,
$$\hat{U}^\dagger \circ \tilde{\op{E}}_\set{G} \circ \hat{U} = \tilde{\op{E}}_\set{G}  \;\;\; \forall\ U \in \set{G}.$$
This symmetrization procedure is referred to as $\set{G}$\textit{-twirling}.  Formally, $\set{G}$-twirling projects the channel onto the invariants of $\set{G}$.
 
For the group $\set{SU}(2^n)$ there are two linearly independent  invariants under twirling.  These may be taken to be the identity channel,  $\hat{I}(\rho) = \rho$, and the completely depolarizing channel, $\op{D}(\rho) = \frac{1}{4^n}\sum_{\mu}  P_\mu \rho P_\mu.$  Given an expression for an error channel in terms of Pauli operators,
$$\op{E}(\rho) = \sum_{\mu\nu} x_{\mu\nu} P_\mu \rho P_\nu,$$ 
the corresponding fully twirled channel is 
\begin{align*}
\tilde{\op{E}}_{\set{SU}(2^n)}(\rho) &= (1-p) \rho + \frac{p}{4^n -1} \sum_{\mu\neq I} P_\mu \rho P_\mu \\
&=\lambda\rho+ \frac{1-\lambda}{4^n} \op{D}(\rho),
\end{align*} 
where $p =\sum_{\mu \neq I} x_{\mu\mu}$ is the entanglement infidelity and
$$\lambda =  1-p\frac{4^n}{4^n-1}$$
is the eigenvalue of the channel for non-identity Pauli operators, i.e., $\tilde{\op{E}}_{\set{SU}(2^n)}(P_\mu) = \lambda P_\mu$ for all $P_\mu\neq I$.  Thus, twirling by $\set{SU}(2^n)$ is sufficient to convert any error channel to one specified by a single real number.

Twirling by $\set{SU}(2^n)$ is not practical experimentally, but neither is it necessary.  It suffices to twirl by any group $\set{G}$ satisfying the following condition: for any polynomial that is second order and homogeneous in both the matrix elements of $U$ and $U^\dagger$ (in the fundamental representation), the expectation value is the same whether $U$ is sampled uniformly from $\set{SU}(2^n)$ or $\set{G}$.  Groups with this property are referred to as 2 designs~\cite{CLLW16}.  A 2 design for $\set{SU}(2^n)$ can equivalently be defined as a group $\set{G}$ satisfying
$$ \ave{a_{\mu\nu}(U)} = 0 \;\;\; \text{and}$$
$$ \ave{\left(a_{\mu\nu}(U)\right)^2} = \frac{1}{4^n-1}$$
for all $\mu,\nu\neq I$ where the overline denotes an average over $U$ and 
\begin{align*}
UP_\mu U^\dagger = \sum_\nu a_{\mu\nu}(U) P_\nu.
\end{align*}
In words
this means that  each non-identity Pauli operator is mapped to every non-identity Pauli operator with average amplitude of zero and equal average square amplitude.  It is straightforward to see that the Clifford group satisfies these conditions, as well as certain subgroups thereof, such as, in the case of a single qubit, the group generated by $e^{i\pi/(3\sqrt{3 })(X+Y+Z)}$ and the $X$ gate.
\iffalse
 equivalently to Since $\tilde{\op{E}}_\set{G}$ is an expectation value of a (2,2)-polynomial in the matrix elements of $U,$ it follows that if $\set{G}$ is a $2$ design
with respect to the group  $\set{SU}(2^n)$, then $$\tilde{\op{E}}_\set{G} = \mathbb{E}_{U\sim \set{G}}  \left[U^\dagger \circ \op{E} \circ U  \right] = \mathbb{E}_{U\sim \set{SU}(2^n)}  \left[U^\dagger \circ \op{E} \circ U \right].$$ \fi 
\iffalse
Thus, twirling an error channel by a unitary $2$ design is equivalent to twirling by the full unitary group where the unitary gates are drawn uniformly (according to the Haar measure). 
The twirling transformation,
$$\tilde{\op{E}}_{\set{G}} = \op{T}_{\set{G}} (\op{E}),$$
is not a unique function of $\set{G}$. Specifically, $\op{T}_{\set{G}'}= \op{T}_{\set{G}}$ for a subset $\set{G}' \subset \set{G}$  when the expectation value of every  (2,2)-polynomial in the matrix elements (in the fundamental representation) of $U$ are the same whether sampling uniformly from $\set{G}'$ or $\set{G}$.
Such a subset $\set{G}'$ is referred to a 2 design with respect to $\set{G}$.  For any 2 design  with respect to the group $\set{SU}(2^n),$ the above definition is equivalent to the following conditions,
$$ \mathbb{E}_{ U \sim \set{G}'}  \left[ \mbox{tr}(U   P_\mu U^\dagger P_\nu )\right] = 0,$$
where $P_\mu$ is a multi-qubit Pauli operator and $P_\mu \neq I.$
\fi 
Twirling by 2 designs such as these forms the basis of standard randomized benchmarking.

\subsection{Standard Randomized Benchmarking Protocol}
Given the gate error channel,
\begin{equation*}
\op{E}(\rho) = \sum_{\mu\nu} x_{\mu\nu} P_\mu \rho P_\nu,
\end{equation*}
the goal of randomized benchmarking is to determine the entanglement infidelity,
\begin{equation*}
p = \sum_{\mu\ne I} x_{\mu\mu},
\end{equation*}  
or some simple function thereof\footnote{In fact, randomized benchmarking papers often quote the average infidelity, which is related to the entanglement infidelity by $p_{\text{ave}}=2^n p/(2^n+1)$~\cite{Nie02}.  We focus on the entanglement infidelity instead since it is more relevant for multi-qubit states and corresponds to the error parameter typically used in simulations of quantum error correction.}.  

The standard randomized benchmarking protocol~\cite{MGE11} consists of many repetitions of the following experiment:
\begin{itemize}
\item{Prepare an initial state, $\rho_0$, such that $\rho_0$ is the $+1$ eigenstate of a projector, $\proj$.}
\item{Perform a random gate sequence,  $U_{l-1}\ldots U_1$,  where each gate is selected independently from  a unitary 2 design.}
\item{Perform one final unitary gate chosen such that in the absence of errors the unitary gate sequence performs the identity\footnote{It is worth mentioning that in practice it's better to return to the starting eigenbasis rather than the starting state.},\
$$U_l = \prod_{t=1}^{l-1} U_{t}^{-1}.$$  }
\item{Measure $\proj$.}
\end{itemize}

Averaged over many runs, the measurement statistics yield a fidelity for the experiment which depends only on the twirled error channel and the preparation and measurement errors.
In order to isolate state preparation and measurement errors, $\op{E}_p$ and $\op{E}_m$, respectively, the length, $l$, of the gate sequences is varied.  The average fidelity for gate sequences of length $l$ is given by
 \begin{eqnarray*}f_l &=& \text{tr}\left( \proj \op{E}_m\left(\tilde{\op{E}}^l(\op{E}_p(\rho_0))\right) \right )\\
&=& \mathbb{E}_{|\vec{U}|=l}( f_\vec{U}), \end{eqnarray*} 
where $f_\vec{U}$ is the fidelity for a given sequence of unitaries $\vec{U} = (U_1,\ldots, U_l)$.

The average sequence fidelities as a function of length are then fit to the decay curve,
\begin{equation*} f_l = c_0 + c_1\lambda^l,
\end{equation*} 
where $\lambda$ is as defined above.

The total number of measurements required is minimized when a different random sequence is selected for each run, but for practical reasons, each gate sequence is typically repeated many times in order to determine the fidelity, $f_\vec{U}$, for each gate sequence, $\vec{U}$.  This also allows additional information about the error channel to be extracted, notably its coherence and non-unitality~\cite{KLR+08}.  The minimum number of gate sequences needed to estimate $f_l$ to a given accuracy depends on the variance of the gate sequence fidelities,
 $$v_l = \sum_{|\vec{U}|=l}(f_\vec{U} - f_l)^2,$$
for which general bounds are derived in the literature~\cite{WF14,HWFW17}.

\section{Clifford Subgroup Twirling\label{sec:CliffordSubgroupTwirling}}
In this  section,  the standard randomized benchmarking protocol is adapted to sampling from subgroups of the Clifford group which are not unitary 2 designs.  All of the groups we consider contain the Pauli group, that is, the group generated by all single-qubit Pauli gates, and so are unitary 1 designs.   It is important to note that twirling with respect to the Pauli group converts an arbitrary channel,  
$$\op{E}(\rho) = \sum_{\mu\nu} x_{\mu\nu} P_\mu \rho P_\nu,$$ 
into the corresponding stochastic Pauli channel, 
 $$\tilde{\op{E}}_\set{P}(\rho) = \sum_{\mu} x_{\mu\mu} P_\mu \rho P_\mu.$$

Each Pauli operator is an eigenoperator of every Pauli channel.  That is, $\tilde{\op{E}}_\set{P} (P_\mu) = \lambda_\mu P_\mu$.  The corresponding eigenvalue is given by,
$$\lambda_\mu\ = \sum_{\nu| [P_\nu,P_\mu]=0} x_{\nu\nu} -  \sum_{\nu| [P_\nu,P_\mu] \neq 0} x_{\nu\nu}.$$

For any subgroup of the Clifford group, $\set{S}$, the orbit of each Pauli operator under the action of $\set{S}$ forms one of a set of $k$ blocks, $\{{\bf B}_0, \ldots, {\bf B}_{k-1}\},$ each containing $N_i(n)$ Pauli operators.
Twirling with respect to a subgroup of the Clifford group which also contains the Pauli group, therefore results in a channel of the form, 
\begin{equation*}
\tilde{\op{E}}_\set{S} (\rho) =  (1-p) \rho + \sum_{i=1}^{k-1} \frac{p_i}{N_i(n)} \sum_{\mu \in {\bf B_i }} P_\mu \rho P_\mu,
\end{equation*}
where ${\bf B}_0 = \{I\}$,
$$p_i =  \sum_{\mu \in {\bf B_i }} x_{\mu\mu}, \text{\hspace{1.5em}and\hspace{2em}} p =  \sum_{i=1}^{k-1} p_i.$$

In a case of imperfect mixing such as this, the fidelity decay curve has the form
$$f_l = c_0 + \sum_{i=1}^{k-1} c_i\lambda_i^l$$
where
$$c_i = \tr\left(\proj \op{E}_m \left(\sum_{\nu\in \bf B_i} \tr\left( P_\nu \op{E}_p(\rho_0)\right) P_\nu\right)\right),$$
$\rho_0$ is the ideal initial state, and $\proj$ is the projector of interest.  Notably, in the multi-qubit case, the final multi-qubit measurement is typically implemented via many single-qubit projective measurements.  Using the measurement results from each qubit separately, many such curves can be extracted concurrently or only a single one of particular interest (e.g., such that only one $c_{i\neq 0}$ is significant).

In theory, the parameters $p_i$ can be determined if it is possible to prepare and measure states in the eigenspace of at least one Pauli operator from each block, either sequentially or all at once using a multi-qubit initial state of the form
$$\rho_0 = \frac{1}{2^n} \prod_\mu \left(I + P_\mu\right).$$
Often, however, preparing an eigenstate for one Pauli operator from each block is impractical in situations where implementing the gate(s) required to convert between blocks is impractical.

Alternatively,  if there exists a Pauli operator, $P$, that commutes with approximately the same fraction of the Pauli operators within each block, then the entanglement fidelity of the error channel can be determined approximately by preparing an eigenstate of $P$ and measuring the decay of the expectation value of the corresponding projector as a function of gate sequence length. This is the primary approach taken in the remainder of the paper.  

\subsection{The Real Clifford Group}
Consider the subgroup of the Clifford group that preserves the evenness or oddness of the number of $Y$ elements in a Pauli string.  This group is referred to as the real Clifford group, and is generated by Hadamard, controlled-NOT, and the single-qubit Pauli gates.    

Twirling with respect to the real Clifford group results in a channel with two non-trivial blocks:
\begin{itemize}
\item{${\bf B}_1$ consists of the non-identity Pauli operators with an even number of $Y$ elements (real Pauli operators).  The size of ${\bf B}_1$ is
$$N_1(n)=\sum_{l\in \text{even}} \binom{n}{l} 3^{n-l} = \frac{4^n + 2^n}{2}-1.$$
}
\item{${\bf B}_2$ consists of the Pauli operators  with an odd number of $Y$ elements (imaginary Pauli operators).  The size of ${\bf B}_2$ is
$$N_2(n) = \sum_{l\in \text{odd}} \binom{n}{l} 3^{n-l}= \frac{4^n - 2^n}{2}.$$
}
\end{itemize}

Given a channel, 
\begin{equation*}
\op{E}(\rho) = \sum_{\nu\mu} x_{\mu\nu}  P_\mu \rho P_\nu,  
\end{equation*}
twirling with respect to the real Clifford group results in the channel,
\begin{align*}
\tilde{\op{E}}_{\set{R}}(\rho)= (1-p) \rho &+\frac{p_1}{N_1(n)} \sum_{\mu \in {\bf B}_1} P_\mu \rho P_\mu  \\
&+\frac{p_2}{N_2(n)} \sum_{\mu \in {\bf B}_2} P_\mu \rho P_\mu,
\end{align*}
where
 $$p_1 =\sum_{\mu \in {\bf B}_1} x_{\mu\mu} \;\;\; \text{and}\;\;\; p_2  = \sum_{\mu \in {\bf B}_2} x_{\mu\mu}. $$

A real Pauli operator anti-commutes with $N_1(n-1) + N_2(n-1)+1$  real Pauli operators and exactly the same number of imaginary Pauli operators.   An imaginary Pauli operator anti-commutes with $2N_1(n-1)+2$ real Pauli operators and $2N_2(n-1)$ imaginary Pauli operators.  It follows that each real Pauli operator is an eigenvector of the twirled channel, $\tilde{\op{E}}_{\set{R}}$, with eigenvalue,
\begin{align*}
\lambda_1 =&1-2 p_1\frac{N_1(n-1) + N_2(n-1)+1}{N_1(n)}\\
&- 2 p_2 \frac{N_1(n-1) + N_2(n-1)+1}{N_2(n)}\\
 =&1- p_1 \frac{4^{n}}{4^n + 2^n-2}-  p_2 \frac{2^{n}}{2^n - 1}\\
=& 1- p_2 - p_1 + \frac{p_1-p_2}{2^{n}} + O(2^{-2n}).
\end{align*} 
while each imaginary Pauli operator is an eigenvector of  $\tilde{\op{E}}_{\set{R}}$  with eigenvalue,
\begin{align*}
 \lambda_2 &=1- 4 p_1\frac{N_1(n-1)+1}{N_1(n)} - 4 p_2\frac{N_2(n-1)}{N_2(n)}\\
 &=1- p_1\frac{2^{n} }{2^n-1} - p_2\frac{4^{n}- 2^{n+1} }{4^n - 2^n}\\
&= 1-p_2 - p_1 + \frac{p_2-p_1}{2^{n}}+ O(2^{-2n}).
\end{align*}

To determine the parameters $p_1$ and $p_2$ it is sufficient to prepare and measure state(s) which are eigenstates of both a real and an imaginary Pauli operator\footnote{Note that any multiqubit state that is an eigenstate of more than one Pauli operator is an eigenstate of at least one non-trivial real Pauli operator.}, but this is typically challenging in cases where only real Clifford gates are available.  In particular, codes for which only the real Clifford gates are transversal generally lack a straightforward procedure for preparation and measurement in the logical $Y$ basis.  Instead, consider the case where the initial state is only an eigenstate of real Pauli operators and therefore only $\lambda_1$ can be extracted.

Recalling that $p=p_1+p_2$, we see that given $\lambda_1$ the entanglement infidelity can be bounded as follows:
$$ \frac{2^n - 1}{2^{n}}(1-\lambda_1) \le p \le \frac{4^n + 2^n-2}{4^{n}} (1-\lambda_1).$$
Using the upper bound as our estimate of $p$ corresponds to assuming that $p_2=0$ and leads to overestimating the entanglement infidelity by a factor of at most $(2^n+2)/2^n$, that is, by a factor of two or less.  For the purpose of benchmarking logical qubits, however, the estimate will typically be much better since logical $Y$ errors are strongly suppressed for many popular codes due to such errors having higher weight and/or separate syndrome measurement and decoding for physical $X$ and $Z$ errors.  In the surface code, for example, twice as many physical errors are required to generate a logical $Y$ error as to generate a logical $X$ or logical $Z$ error.

\subsection{Controlled-NOT and Pauli Gates}
Now consider the subgroup of the Clifford group generated by controlled-NOT and the single-qubit Pauli gates.  Controlled-NOT gates generate the group $\set{GF}_2$ through their action on Pauli operators containing only $X$ and $I$ elements and separately on Pauli operators containing only $Z$ and $I$ elements.
Twirling with respect to this group results in a channel consisting of four blocks:
\begin{align*}
\tilde{\op{E}}_{\set{C}}(\rho) =& (1-p) \rho +\frac{ p_1}{N_1(n)} \sum_{\mu \in {\bf B}_1} P_\mu \rho P_\mu  \\
&+ \frac{p_2}{N_2(n)} \sum_{\mu \in {\bf B}_2} P_\mu \rho P_\mu + \frac{p_3}{N_3(n)} \sum_{\mu \in{\bf B}_3} P_\mu \rho P_\mu\\
&+ \frac{p_4}{N_4(n)} \sum_{\mu \in {\bf B}_4} P_\mu \rho P_\mu,
\end{align*}
where
\begin{itemize}
\item{${\bf B}_1$ consists of the non-identity Pauli operators  containing only $Z$ and $I$ elements.   The size of ${\bf B}_1$ is
$$N_1(n)  = 2^n-1.$$
}
\item{${\bf B}_2$ consists of the non-identity Pauli operators  containing only $X$ and $I$ elements.   The size of ${\bf B}_2$ is
$$ N_2(n) = 2^n -1.$$
}
\item{${\bf B}_3$ consists of the non-identity Pauli operators containing an even number of $Y$ elements and not belonging to ${\bf B}_1$ or ${\bf B}_2$.   The size of ${\bf B}_3$ is
$$N_3(n) = \frac{4^n - 3 \cdot 2^n}{2}    +1.$$ }
\item{${\bf B}_4$ consists of those  Pauli operators containing an odd number of $Y$ elements.   The size of ${\bf B}_4$ is
$$N_4(n) = \frac{4^n - 2^n}{2}.$$
}
\end{itemize}

The eigenvalues of $\tilde{\op{E}}_{\set{C}}$ with respect to the Pauli operators of each block are,
\begin{align*}
\lambda_1 &= 1- (p_2+p_3+p_4)  \frac{2^n}{2^n-1} \\
&= 1- (p_2+p_3+p_4)\left(1+\frac{1}{2^n}\right) + O(2^{-2n}),\\
\lambda_2 &= 1- (p_1+p_3+p_4)  \frac{2^n}{2^n-1} \\
&= 1- (p_1+p_3+p_4)\left(1+\frac{1}{2^n}\right) +O(2^{-2n}),\\
\lambda_3 &= 1- (p_1  + p_2 + p_4)\frac{2^n}{2^n-1} - p_3 \frac{4^{n} - 2^{n+2}}{4^n-3\times 2^n+2}\\
&= 1- p + \frac{p_3-p_1-p_2-p_4}{2^n} + O(2^{-2n}),\\
\lambda_4 &=  1- (p_1+p_2+p_3) \frac{2^{n}}{2^n-1} -  p_4 \frac{2^{n}-2}{2^n-1}  \\
&= 1 - p + \frac{ p_4- p_1- p_2 - p_3}{2^n} + O(2^{-2n}).
\end{align*}

The ability to prepare and measure eigenstates of $X$, $Y$, and $Z$ would enable the reconstruction of all four parameters, $p_1$, $p_2$, $p_3$, and $p_4$.  For logical qubits, at least, preparation and measurement in the $X$ and $Z$ logical bases are often relatively straightforward as most codes of interest are CSS codes.  As discussed in the previous section, however, preparing and measuring in the logical $Y$ basis is often problematic.  The entanglement infidelity averaged over the gates set, $p$, can be estimated for two or more qubits by performing benchmarking on eigenstates of Pauli operators in $\set{B}_1$ and eigenstates of Pauli operators in $\set{B}_2$ independently.
Given $\lambda_1$ and $\lambda_2$, the entanglement fidelity can be bounded as follows:
$$ \frac{2^n - 1}{2^{n+1}}(2-\lambda_1-\lambda_2) \le p \le \frac{2^n - 1}{2^{n}} (2-\lambda_1-\lambda_2).$$
Using the upper bound corresponds to assuming that $p_3, p_4 =0$ and leads to an overestimate of $p$ by at most a factor of $2$ independent of the number of qubits in the benchmarking experiment.  For logical qubits the estimate will typically be much better because $p_3$ and $p_4$ are likely to be much smaller than $p_1$ and $p_2$ for the reasons discussed in the previous section. Alternatively, for $n>2$ preparation and measurement of an eigenstate of a Pauli operator in $\set{B}_3$ (e.g., $\ket{{+}00}$) allows $\lambda_3$ to be extracted from the decay curve.  Given $\lambda_3$, the bounds on $p$ are
$$ \frac{2^n - 1}{2^{n}}(1-\lambda_3) \le p \le \frac{4^n-3\times 2^n+2}{4^{n} - 2^{n+2}} (1-\lambda_3),$$
where the upper bound corresponds to taking $p_3$ to be the only non-zero probability.  This is an unrealistic but conservative assumption, causing us to overestimate $p$ by at most a factor of $(2^n-2)/(2^n-4)$.  For logical qubits the lower bound will typically yield a better estimate of the entanglement infidelity.

\section{Conclusion\label{sec:conclusion}}
In this article we have introduced a method for analyzing the behavior of randomized benchmarking as it applies to subgroups of the Clifford group that do not form 2 designs with respect to $\set{SU}(2^n)$.  We have additionally applied this method to two subgroups of interest.  The first subgroup considered was the real Clifford group, which is generated by controlled-NOT, Hadamard, and the Pauli gates.  We described a protocol for performing randomized benchmarking on $n$ qubits using only the real Clifford group that estimates the entanglement infidelity of the average error channel to within a factor of $(2^n+2)/2^n$.  The second subgroup considered was that generated by controlled-NOT and Pauli gates.  Given the ability to prepare and measure both $\ket{0}$ and $\ket{+}$, we found that the entanglement infidelity can be estimated to within a of factor of either $2$ or $(2^n-2)/(2^n-4)$ depending on which decay constants are extracted.  These results demonstrate that highly accurate approximate randomized benchmarking can be performed without sampling from a unitary 2 design or any approximation thereof.   

\acknowledgements

We thank Michael Mullan for his careful reading of the document.  


%\bibliography{NGCitations}

\end{document}